\documentclass{mn2e}
\usepackage{epsfig}
\usepackage{amsmath}
\def\msun{{\rm M_{\odot}}}

\def\me{{\dot M_{\rm Edd}}}

\def\mo{{\dot M_{\rm out}}}
\def\le{{L_{\rm Edd}}}

\title[Large--Scale Outflows in Galaxies]{Large--Scale Outflows in Galaxies}

\author[A.R. King, K. Zubovas, C.Power] {\parbox{18cm}{A.R. King$^{1}$,
    K. Zubovas$^{1}$, C. Power$^{1}$}\\ 
$^1$Theoretical Astrophysics
  Group, University of Leicester, Leicester LE1 7RH}

\date{\today}

\volume{000}

\pagerange{\pageref{firstpage}--\pageref{lastpage}} \pubyear{2005}

\begin{document}

\label{firstpage}

\maketitle

\begin{abstract}
We discuss massive outflows in galaxy bulges, particularly ones driven
by accretion episodes where the central supermassive black hole
reaches the Eddington limit. We show that the quasar radiation field
Compton--cools the wind shock until this reaches distances $\sim
1$~kpc from the black hole, but becomes too dilute to do this at
larger radii. Radiative processes cannot cool the shocked gas within
the flow time at any radius. Outflows are therefore momentum--driven
at small radii (as required to explain the $M - \sigma$ relation). At
large radii they are energy--driven, contrary to recent claims.

We solve analytically the motion of an energy--driven shell after the
central source has turned off. This shows that the thermal energy in
the shocked wind can drive further expansion for a time $\sim 10$
times longer than the active time of the central source. Outflows
observed at large radii with no active central source probably result
from an earlier short (few Myr) active phase of this source.
\end{abstract}

\renewcommand{\thefootnote}{\fnsymbol{footnote}}
\footnotetext[1]{E-mail: {\tt andrew.king@astro.le.ac.uk }}

\section{Introduction}

The large--scale structure of galaxies often has surprisingly close
connections to properties of their nuclei. The $M - \sigma$ relation
between supermassive black hole (SMBH) mass $M$ and bulge velocity
dispersion $\sigma$ is the most striking of these. Similar relations
hold between black hole and galaxy stellar bulge mass $M_b$, and
between the mass of nuclear star clusters and $\sigma$ in galaxies
where there is no strong evidence for the presence
of an SMBH ($\sigma \la 150~{\rm km\,s^{-1}}$).

Massive gas outflows driven by the central object offer a way of
connecting these apparently disparate scales. A fast wind from the
nucleus collides with the host galaxy's interstellar medium, driving a
reverse shock into the wind, and a forward shock into the ISM. This
shock pattern moves outwards at a speed mainly determined by whether
or not the reverse shock cools on a time short compared with the
outflow timescale ($R/\dot R$) or not (cf Dyson \& Williams, 1997;
Lamers \& Cassinelli, 1997). In the first case (efficient cooling),
only the ram pressure of the original outflow is communicated to the
ambient medium. This is a {\it momentum--driven} flow. In the second
case (inefficient cooling) the full energy of the fast wind is
communicated to the ambient medium through its thermal expansion after
the shock. This is an {\it energy--driven} flow, which expands at
higher speed and so can have a much larger effect on the bulge of the
host galaxy.

Both types of outflow are important in galaxy formation. This paper is
mainly concerned with the large--scale effects of energy--driven
flows. The existence of flows of this type has recently been
questioned, so we first set the problem in context. We show that
energy--driven outflows do occur, and are ubiquitous on large
scales. Solving the outflow equations analytically, we give a simple
relation between the time that the outflow is driven by the central
source, and the time over which it can be observed as coasting after
this source turns off. This relation means that observed outflows can
be used to constrain the past activity of a source. In this paper we
deal with the case where this source is a quasar, which we model as an
Eddington--accreting SMBH. Similar considerations apply in cases where
the driving source is a nuclear star cluster.

\section{Momentum or Energy Driving?}

The first proposal that outflows might relate SMBH and galaxy
properties was by Silk \& Rees (1998, hereafter SR98), who considered
the effect of an Eddington wind from the black hole colliding with the
host ISM. Requiring the shock pattern to move with the escape
velocity, and so presumably cutting off accretion to the black hole,
they found $M \propto \sigma^5$, with an undetermined coefficient of
proportionality. Later, King (2003; 2005) pointed out that SR98
implicitly assumed an energy--driven outflow, whereas Compton cooling in
the radiation field of the active nucleus was likely to produce a
momentum--driven flow. The condition that this flow should be able to
escape the immediate vicinity of the black hole, and so cut off
accretion, predicts a black hole mass
\begin{equation}
M = {f_g\kappa\over \pi G^2}\sigma^4 \simeq 2\times
10^8\msun\sigma_{200}^4
\label{msig}
\end{equation}
in good agreement with the observed relation (Ferrarese \& Merritt,
2000; Gebhardt et al., 2000) (which is itself probably an upper limit
to the SMBH mass, cf Batcheldor, 2010; King, 2010b). Here $f_g \simeq
0.16$ is the gas fraction, $\kappa$ the electron scattering opacity,
and $\sigma_{200}$ the velocity dispersion in units of
200~km\,s$^{-1}$. By contrast, an energy--driven outflow as in SR98
would have produced a mass smaller than (\ref{msig}) by a factor $\sim
\sigma/c \sim 10^{-3}$ (e.g. King, 2010a). A later application of
similar ideas (McLaughlin et al., 2006, Nayakshin et al, 2009) to
outflows driven by nuclear star clusters shows that these produce an
offset $M - \sigma$ relation between total cluster mass and velocity
dispersion, the offset resulting from the fact that star clusters
produce roughly 20 times less outflow momentum per unit mass compared
with an accreting black hole.

There have also been attempts to explain the relation between black
hole and bulge stellar mass in terms of Eddington outflows from
accreting SMBH. The observed relation $M \sim 10^{-3}M_b$ (cf H\"aring
\& Rix, 2004) means that this an inherently more complex problem than
$M - \sigma$, since $M_b$ is apparently the small part that remains
after some process has almost swept the bulge clear of its orginal
baryon content. Two recent papers discuss this problem. 

Power et al. (2011) suggest that star formation in a galaxy bulge is
self--limiting, and this limit largely determines the bulge stellar
mass $M_b$. They further suggest that an energy--driven outflow from
the central black hole clears away the remaining gas. This process
cannot be totally effective: King (2010b) shows that energy--driven
outflows are Rayleigh--Taylor unstable since the rapid expansion of
the shocked wind leads to a large density contrast with the ambient
medium. Thus a fraction of the gas can still remain even after the
outflow passes.

In contrast, Silk \& Nusser (2010) assert that energy--driven outflows
do not occur at all in galaxy bulges. It is easy to show that
momentum--driven outflows cannot clear the remaining gas from the
bulge (Silk \& Nusser, 2010; Power et al., 2011, Appendix).
Accordingly Silk \& Nusser (2010) suggest that star formation must be
able to remove it.

\begin{figure}
  \centering
    \includegraphics[width=0.5\textwidth]{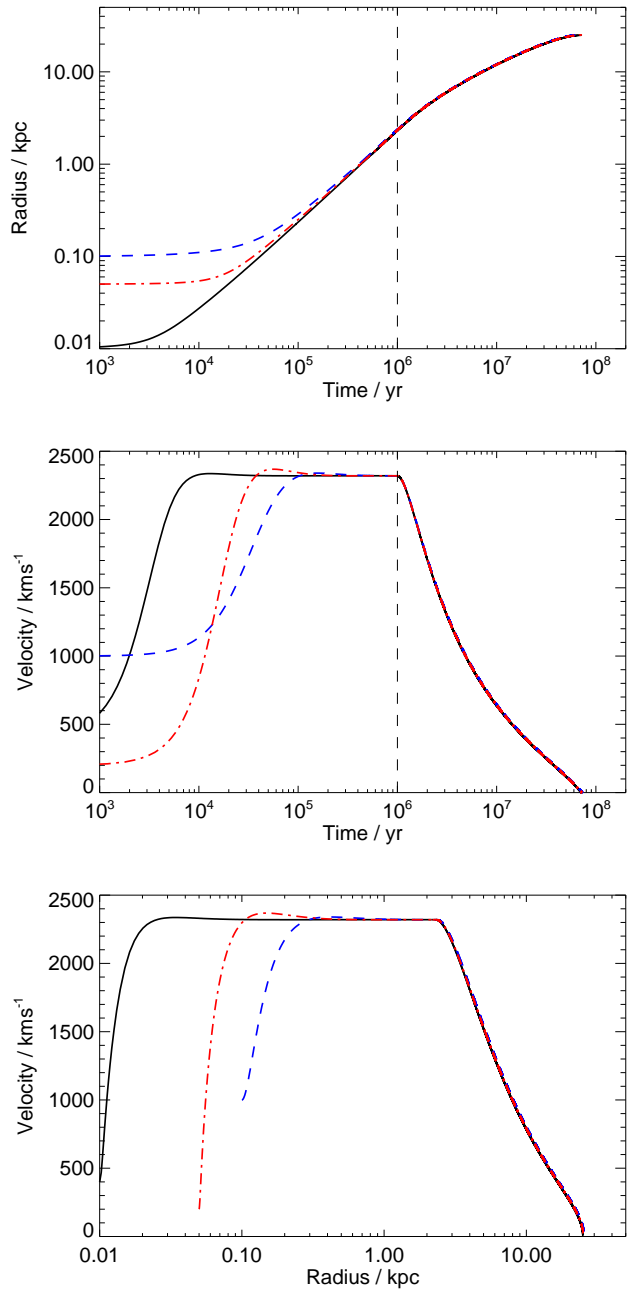}
    \caption{Evolution of an energy--driven shock pattern for the case
      $\sigma = 200$ kms$^{-1}, f_g = 10^{-2}$ computed numerically
      from the full equation (\ref{motion}). Top: radius vs time,
      middle: velocity vs time, bottom: velocity vs radius. The curves
      refer to different initial conditions: black solid -- $R_0 = 10$
      pc, $v_0 = 400$ km/s; blue dashed -- $R_0 = 100$ pc, $v_0 =
      1000$ km/s; red dot-dashed - $R_0 = 50$ pc, $v_0 = 200$
      km/s. All these solutions converge to the attractor
      (\ref{ve}). The vertical dashed line marks the time $t = 10^6$
      yr when the quasar driving is switched off.  All solutions then
      follow the analytic solution (\ref{dotr}).}
    \label{varIC}
\end{figure}

\begin{figure}
  \centering
    \includegraphics[width=0.5\textwidth]{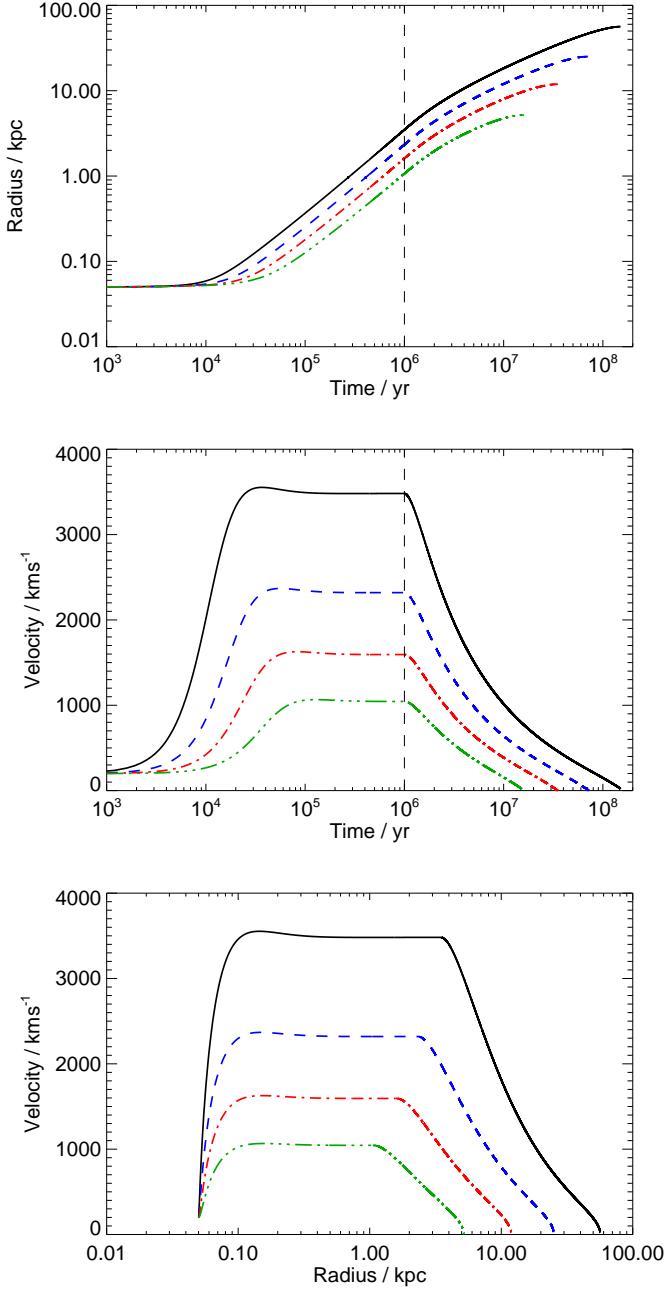}
    \caption{Same as Figure 1, but with $R_0 = 50$ pc and $v_0 = 200$ km/s
and varying mean gas fractions: black solid - $f_g = 3 \cdot 10^{-3}$;
blue dashed - $f_g = 10^{-2}$; red dot-dashed - $f_g = 3 \cdot 10^{-2}$;
green dash-triple-dotted - $f_g = 10^{-1}$.}
    \label{varfg}
\end{figure}

\section{Shock Cooling}

To decide whether energy--driven outflows exist or not we consider an
Eddington wind ($\mo \simeq \me$) from a supermassive black hole
propagating in an approximately isothermal galaxy bulge, with gas
density
\begin{equation}
\rho = {f_g\sigma^2\over 2\pi Gr^2}.
\label{rho}
\end{equation}
The gas mass inside radius $R$ is
\begin{equation}
M(R) = 4\pi\int_0^R\rho r^2 {\rm d}r = {2f_g\sigma^2R\over G}.
\label{m}
\end{equation}
As we have seen, the important question for the gas motions is whether
the reverse shock cools. The preshock wind has a velocity $v \simeq
\eta c \simeq 0.1c$ (King \& Pounds, 2003; King, 2010a), which implies
a (reverse) shock temperature
\begin{equation}
T_s \simeq {3\over 16}{\mu m_H\over k}v^2 \simeq 1.6\times 10^{10}~{\rm K}.
\label{tshock}
\end{equation}
This gas is clearly too hot to have any bound electrons, so the only
losses cooling it are Compton and free-free. The mass conservation
equation for the Eddington outflow gives a postshock number density
\begin{equation}
N = 4\times{\mo\over 4\pi R^2\mu m_Hv} \simeq 1 \times
10^{-3}(\mo/\me)M_8R_{\rm kpc}^{-2}~{\rm cm^{-3}}
\end{equation}
where $M_8$ is the SMBH mass in units of $10^8\msun$ and $R_{\rm kpc}$
is the radial distance in kpc. This gives a radiative (free--free)
cooling time for the shocked gas of
\begin{equation}
t_{\rm rad} \simeq 2\times 10^{11}M_8^{-1}R_{\rm kpc}^2~{\rm yr}.
\label{ff}
\end{equation}
King (2003, eqn 8) shows that the Compton cooling time of this gas in the
quasar radiation field is
\begin{equation}
t_c = {2\over 3}{cR^2\over GM}\biggl({m_e\over m_p}
\biggr)^2\biggl({c\over v}\biggr)^2b
\simeq 10^7R_{\rm kpc}^2bM_8^{-1}~{\rm yr}
\label{tc}
\end{equation}
where $b\sim 1$ is the fractional solid angle of
the outflow, $m_e, m_p$ are the electron and proton masses, and we
have set $v = 0.1c$ in the original equation.

To decide if cooling is effective we compare these timescales with the
flow timescale for a momentum--driven outflow, which is
\begin{equation}
t_{\rm flow} = \frac{R}{\dot{R}} = 5\times 10^6R_{\rm kpc}\sigma_{200}M_8^{-1/2}~{\rm yr},
\end{equation}
(cf. King 2003, eqns 9 \& 14). We find directly
\begin{equation}
{t_c\over t_{\rm flow}}  = 
1.8R_{\rm kpc}\sigma_{200}^{-1}M_8^{-1/2}b.
\end{equation}

We see that Compton cooling is effective only out to about $R =
1$~kpc, (cf Ciotti \& Ostriker, 1997) while the radiative (free--free)
cooling is always far longer than the flow time. Silk \& Nusser (2010)
claim the opposite, but appear to have considered the cooling of the
ambient gas rather than the shocked wind which contains all the
energy. Their adopted cooling function (Sutherland \& Dopita, 1993)
only goes to temperatures $10^7 - 10^8$~K, far below the shock
temperature $T_s \simeq 10^{10}$~K. We recover the result (King, 2003;
2005) that in a galaxy bulge an Eddington outflow is momentum--driven
when very close to the SMBH, but becomes energy--driven outside a
typical radius $\sim 1$~kpc.

Many galaxies show evidence for massive high--speed ($v \sim 1000~{\rm
  km\, s^{-1}}$) gas outflows on large scales ($\sim 20$~kpc)
(e.g. Tremonti et al., 2007; Holt et al. 2008). By the arguments of
this Section, these must be energy--driven.  Their ultimate cause may
be starbursts, or AGN activity by the central SMBH.  However these
nuclear phenomena are often absent or weak when the outflows are
observed. So to understand the connection between the observed outflow
and its original cause we need to know how the outflow coasts and
ultimately stalls in the absence of driving.

\section{Energy--Driven Outflows}

The equation governing the movement of the shock pattern in an
energy--driven outflow in an isothermal potential is (King, 2005)
\begin{equation}
{\eta\over 2}\le = 
{2f_g\sigma^2\over G}\biggl\{ {1\over 2}R^2\dddot R + 3R\dot R\ddot R
+ {3\over 2}\dot R^3\biggr\} + 10f_g{\sigma^4\over G}\dot R
\label{en}
\end{equation}
with $\eta \simeq 0.1$ the accretion efficiency, $\le$ the Eddington
luminosity of the central black hole, $\sigma$ the velocity dispersion
of the ambient medium and $f_g$ the gas fraction relative to all
matter in this medium. The latter quantity may be depleted relative to
its value $f_c$ prevailing when the earlier momentum--driven outflow
establishes the $M - \sigma$ relation (\ref{msig}). Using the expression
$M = \frac{f_g \kappa}{\pi G^2} \sigma^4$ in $\le$ gives
\begin{equation}
\eta c \sigma^2{f_c\over f_g} =\biggl\{ {1\over 2}R^2\dddot R + 3R\dot R\ddot R
+ {3\over 2}\dot R^3 \biggr\} + 5\sigma^2 \dot R
\label{motion}
\end{equation}
This equation has a solution of the form $R = v_et$ with
\begin{equation}
2\eta c {f_c\over f_g} = 3{v_e^3\over \sigma^2} + 10v_e
\end{equation}
[Note that in King (2005), which considered the case $f'_g = f_g$, a
  factor 2 was omitted from the lhs of the corresponding equation
  (19). The subsequent algebra is nevertheless correct.] The assumption $v_e <<
\sigma$ leads to a contradiction ($v_e \simeq 0.02c[f_c/f_g] >>
\sigma$), so the equation has the approximate solution
\begin{equation}
v_e \simeq \biggl[{2\eta f_c\over 3f_g}\sigma^2c\biggr]^{1/3} \simeq
925\sigma_{200}^{2/3}(f_c/f_g)^{1/3}~{\rm km\ s}^{-1}
\label{ve}
\end{equation} 

This solution is an attractor. At radii $R$ large enough that Compton
cooling becomes ineffective, the extra gas pressure makes the 
previously momentum--driven shock
pattern accelerate to this value. 

At still larger radii, it may happen that the quasar supplying the
driving term on the lhs of equation (\ref{motion}) switches
off. Evidently the shock pattern will continue to propagate outwards
for a time, because of the residual gas pressure in the shocked
wind. Its equation of motion now becomes
\begin{equation}
{1\over 2}R^2\dddot R + 3R\dot R\ddot R + {3\over 2}\dot R^3 +
5\sigma^2 \dot R = 0
\label{coast}
\end{equation}
As the independent variable $t$ does not appear in this equation, we
let $\dot R = p$, and replace $\ddot R = pp', \dddot R = p^2p'' +
pp'^2$, where the primes denote differentiation wrt $R$. After a
little algebra, the equation takes the form
\begin{equation}
{R^2\over 2}{{\rm d}\over {\rm d}R}(pp') + 3Rpp' + {3\over 2}p^2 +
  5\sigma^2 = 0
\end{equation}
or
\begin{equation}
{1\over 4}R^2y'' + {3\over 2}Ry' + {3\over 2}y + 5\sigma^2 = 0,
\end{equation}
where $y = p^2$.
Now we write $y = y_1 - 10\sigma^2/3$ to reduce the equation to
the algebraically homogeneous form
\begin{equation}
R^2y_1'' + 6Ry_1' + 6y_1 = 0
\end{equation}
which has linearly independent solutions $y_1 \propto R^{-2},
R^{-3}$. Reversing the earlier substitutions we have 
\begin{equation}
p^2 = \dot R^2 = {A_2\over R^{2}} + {A_3\over R^{3}} - {10\over 3}\sigma^2
\end{equation}
We now choose the constants $A_2, A_3$ to fulfil the boundary
conditions $\ddot R = 0, \dot R = v_e$ at the shock position $R = R_0$
where the quasar turns off. This gives finally
\begin{equation}
\dot R^2 = 3\biggl(v_e^2 + {10\over 3}\sigma^2\biggr)\biggl({1\over
  x^2} - {2\over 3x^3}\biggr) - {10\over 3}\sigma^2
\label{dotr}
\end{equation}
where $x = R/R_0\geq 1$. Figure (1) shows numerical solutions of the
full equation of motion. With an arbitrary initial condition at small
$R$, the shock pattern rapidly adopts the constant velocity
$v_e$. Once the quasar switches off, the velocity decays as predicted
by the exact solution (\ref{dotr}).

Equation (\ref{dotr}) gives the velocity of the shock pattern after
the quasar switches off. This pattern stalls (i.e. $\dot R = 0$) when
\begin{equation}
{1\over x^2} - {2\over 3x^3} = {10\sigma^2\over 9(v_e^2 +
  10\sigma^2/3)}.
\label{stall}
\end{equation}
Since $v_e >> \sigma$ we must have $x >> 1$, so we can neglect the
$1/x^3$ term on the rhs of (\ref{stall}) to get
\begin{equation}
x_{\rm stall}^2 \simeq {9\over
  10}\biggl({v_e^2\over \sigma^2} + {10\over 3}\biggr) \simeq
{9v_e^2\over 10\sigma^2}
\end{equation}
where we have used $v_e >> \sigma$ at the last step. So finally
\begin{equation}
R_{\rm stall} \simeq 0.95{v_e\over \sigma} R_0 \simeq 0.95\biggl[{2\eta
    f_c c\over
    3f_g\sigma}\biggr]^{2/3} R_0
\label{rstall} 
\end{equation}

We can find a good approximation for the delay between quasar turnoff and
the shock stalling by integrating eq. (\ref{dotr}). Again neglecting
the $1/x^3$ term this reduces to a quadrature of the form
\begin{equation}
t = \int_{R_0}^{(C/D)^{1/2}} {R{\rm d}R\over (C - DR^2)^{1/2}}
\end{equation}
with 
\begin{equation}
C = 3\biggl(v_e^2 + {10\over 3}\sigma^2\biggr)R_0^2,\ \ \ D = {10\over
  3}\sigma^2
\end{equation}
We find
\begin{equation}
t \simeq \frac{(C-DR_0^2)^{1/2}}{D} \simeq {R_0v_e\over 2\sigma^2} \simeq
{R_{\rm stall}\over 2\sigma}
\end{equation}

The shock pattern moves at the speed $v_e$ for almost all the time
that the quasar is on, so we can write
\begin{equation}
R_0 \simeq v_et_{\rm acc}
\end{equation}
where $t_{\rm acc}$ is the timescale over which the central black hole
accretes at the Eddington rate. Using (\ref{rstall}) we can rewrite
this as
\begin{equation}
R_{\rm stall} \simeq {v_e^2\over \sigma}t_{\rm acc}
\label{rstall2}
\end{equation}
which of course implies
\begin{equation}
t_{\rm stall} \simeq \biggl({v_e\over \sigma}\biggr)^2{t_{\rm
    acc}\over 2}.
\end{equation}

This last relation is interesting, because it shows that outflows
persist for quite a long time after the quasar switches off. Using
(\ref{ve}) we find
\begin{equation}
t_{\rm stall} \simeq 10t_{\rm acc}\sigma_{200}^{-2/3}(f_c/
  f_g)^{2/3} 
\end{equation}
Hence outflows can in principle persist for an order of magnitude
longer than the driving phases giving rise to them.

\section{Escape}

We can use the results of the last Section to find the conditions for
SMBH growth to remove gas from the host galaxy bulge. Attempts to
explain the relation between SMBH and bulge mass (e.g. King, 2003;
2005; Silk \& Nusser, 2010) often invoke this kind of process. A
complication so far not treated is that energy--driven outflows are
Rayleigh--Taylor unstable, and the bulge mass remaining may depend on
the nonlinear growth of these instabilities. Nevertheless it seems
probable that significant mass removal requires much of the shocked
gas to escape the galaxy.

This happens if the shock pattern reaches the galaxy's virial radius
\begin{equation}
R_V \simeq {\sigma\over 7H} = {\sigma t_H\over 7h(z)}
\label{rvirial}
\end{equation}
before stalling. Here $H = H_0h(z)$, with $H_0$ the Hubble constant,
and $h(z)$ gives the redshift dependence. Requiring $R_{\rm stall} >
R_V$ and using (\ref{rstall2}) gives
\begin{equation}
t_{\rm acc} > 1\times 10^8\biggl({\eta_{0.1}f_g\over
  f_c}\biggr)^{2/3}\sigma_{200}^{2/3}~{\rm yr}
\end{equation}
where $\eta_{0.1} = \eta/0.1$. This is about twice the Salpeter
timescale for the mass growth of the SMBH, almost independently of
other parameters. Apparently the black hole must grow significantly in
order to remove a significant amount of bulge mass. 

We may compare this accretion timescale with the time required for the SMBH
luminosity to unbind the gas in the galaxy. Using eq. (\ref{m}) with $R=R_V$
from eq. (\ref{rvirial}), assuming that the gas binding energy is $E_b 
\sim M \sigma^2$ and the SMBH energy input $E_{BH} = 0.05 \xi_5 L_E t_{\rm vir}$ 
(typical for an energy-driven outflow) gives
\begin{equation}
t_{\rm vir} \sim \frac{f_g}{f_c} \frac{M_\sigma}{M} \frac{R_V}{2\xi c} \simeq
  1.5 \times 10^7 \frac{f_g}{f_c} \xi_5^{-1}~{\rm yr}.
\end{equation}

We see that the time it takes for an Eddington-limited accreting SMBH to
inject enough energy into the gas to unbind it is, in principle, shorter
than the Salpeter time. However, crucially, this luminosity has to be
communicated to the gas in the host galaxy. Communication via an energy--driven
wind therefore requires an accretion timescale $t_{\rm acc}$ due to the wind outflow
having $v_e \ll c$.

If the galaxy is inside a cluster, the outflow may reheat the cluster gas
(King 2009). In this case, the virial radius of a galaxy is not well defined,
but we may consider how long it takes for an outflow from the central cluster
galaxy to reach the typical cluster cooling core radius $R_{\rm core} \simeq 150 
\sigma_{1000}^{1/2}$ kpc, where $\sigma_{1000}$ is the cluster velocity dispersion 
in units of $1000$ km/s. If the galaxy has $\sigma = 200$ km/s and $f_g \sim f_c$,
then the outflow cannot propagate into the intracluster medium, as $v_e \simeq
\sigma_c$. However, if we take the velocity dispersion of the surrounding material
to be similar to that in a galaxy, then the accretion duration is

\begin{equation}
t_{\rm acc,c} \simeq 3.4 \times 10^7 \sigma_{200}^{-1/3} 
   \left(\frac{f_c}{f_g}\right)^{-2/3} \; \mathrm{yr},
\end{equation}
and the stalling time is $t_{\rm stall,c} \simeq 3.7 \times 10^8 \sigma_{200}^{-1}$ yr.
This is the timescale on which the intracluster medium is replenished by the
outflow from the central galaxy, provided that the outflow occurs. As long as
the AGN duty cycle of the SMBH at the centre of that galaxy is greater than
$f \ge t_{\rm stall,c}/t_{\rm H} \simeq 2.7 \%$, the intracluster medium is continuously 
replenished and reheated, as the temperature of the gas in the snowplough 
phase (the outer shock) of the outflow is $T_{out} \sim 10^8$ K, similar to 
the virial temperature of the cluster gas.

\section{Visibility}

The most favourable
case for viewing outflows is when each quasar phase is sufficiently
short that the associated outflow has not left the visible galaxy by
the time it stalls. If for example we take the visible galaxy to have
a size $\sim 20$~kpc, we want $R_{\rm stall} \la 20$~kpc, which by 
(\ref{rstall2}) requires 
\begin{equation}
t_{\rm acc} \la 1.7\times 10^6\sigma_{200}^{-4/3}(f_c/f_g)^{-2/3}~{\rm
  yr}.
\label{acc}
\end{equation}
Thus short growth episodes like this are most favourable for seeing
outflows. The fraction of galaxies actually showing outflows then
depends on the growth time of their black holes. The frequency of
detectable outflows in principle offers a way of constraining
the growth history of supermassive black holes. 

\section{Discussion}

This paper has discussed massive outflows in galaxy bulges, chiefly
those driven by accretion episodes where the central supermassive
black hole reaches the Eddington limit. We have shown that these
outflows are momentum--driven at sizes $R \la 1$~kpc, as required to
explain the $M - \sigma$ relation, but become energy--driven at larger
radii because the quasar radiation field becomes too dilute to cool
the wind shock within the flow time. Radiative cooling is incapable of
doing this in any regime, contrary to recent claims.
  
We derive an analytic solution of the equation governing the motion of 
an energy--driven shell after the central source has turned off. We
show that the thermal energy in the shocked wind is able to drive
further expansion for a time typical 10 times longer than the original
driving time. Outflows observed at large radii with no active central
source probably result from an earlier short (few Myr) active phase of
this source. 

Energy--driven outflows from longer--lasting accretion episodes escape
the galaxy, and may well be responsible for removing ambient gas from
the bulge, as required in some pictures of the black hole -- bulge
stellar mass relation. We stress however that since these outflows are
Rayleigh--Taylor unstable, some gas make leak through the shocks and
not be swept out. This problem is impossible to handle analytically
and is currently numerically intractable. The inherent difficulty is
that the instability sets in at very short wavelengths, placing great
demands on spatial resolution.

\section{Acknowledgments}

KZ acknowledges an STFC studentship and CP an STFC postdoctoral
position. Theoretical astrophysics research in Leicester is supported
by an STFC rolling grant.


\begin{thebibliography}{}

\bibitem[\protect\citeauthoryear{Batcheldor}{2010}]{2010ApJ...711L.108B} 
Batcheldor D., 2010, ApJ, 711, L108 

\bibitem[\protect\citeauthoryear{Ciotti \&
    Ostriker}{1997}]{1997ApJ...487L.105C} Ciotti L., Ostriker J.~P.,
  1997, ApJ, 487, L105

\bibitem{} Dyson, J.E., Williams, D.A., 1997, {\it The Physics of the
  Interstellar Medium}, Institute of Physics Publishing, Bristol and
  Philadelphia

\bibitem[\protect\citeauthoryear{Ferrarese \&
    Merritt}{2000}]{2000ApJ...539L...9F} Ferrarese L., Merritt D.,
  2000, ApJ, 539, L9

\bibitem[\protect\citeauthoryear{Gebhardt et 
al.}{2000}]{2000ApJ...539L..13G} Gebhardt K., et al., 2000, ApJ, 539, L13 

\bibitem[\protect\citeauthoryear{H{\"a}ring \&
    Rix}{2004}]{2004ApJ...604L..89H} H{\"a}ring N., Rix H.-W., 2004,
  ApJ, 604, L89

\bibitem[\protect\citeauthoryear{Holt, Tadhunter, \&
    Morganti}{2008}]{2008MNRAS.387..639H} Holt J., Tadhunter C.~N.,
  Morganti R., 2008, MNRAS, 387, 639

\bibitem[\protect\citeauthoryear{King}{2003}]{2003ApJ...596L..27K} King A.~R., 
2003, ApJ, 596, L27 

\bibitem[\protect\citeauthoryear{King}{2005}]{2005ApJ...635L.121K} King A.~R., 
2005, ApJ, 635, L121 

\bibitem[\protect\citeauthoryear{King}{2009}]{2009ApJ...695L.107K} King A.~R., 
2009, ApJ, 695, L107 


\bibitem[\protect\citeauthoryear{King}{2010a}]{2010MNRAS.402.1516K} King 
A.~R., 2010a, MNRAS, 402, 1516 

\bibitem[\protect\citeauthoryear{King}{2010b}]{2010MNRAS.408L..95K} King 
A.~R., 2010b, MNRAS, 408, L95 


\bibitem[\protect\citeauthoryear{King \&
    Pounds}{2003}]{2003MNRAS.345..657K} King A.~R., Pounds K.~A.,
  2003, MNRAS, 345, 657


\bibitem{}Lamers, H.J.G.L.M., Cassinelli, J.P., 1997, {\it
  Introduction to Stellar Winds}, Cambridge University Press,
  Cambridge U.K.

\bibitem[\protect\citeauthoryear{McLaughlin, King, \&
    Nayakshin}{2006}]{2006ApJ...650L..37M} McLaughlin D.~E., King
  A.~R., Nayakshin S., 2006, ApJ, 650, L37


\bibitem[\protect\citeauthoryear{Nayakshin, Wilkinson, \&
    King}{2009}]{2009MNRAS.398L..54N} Nayakshin S., Wilkinson M.~I.,
  King A., 2009, MNRAS, 398, L54


\bibitem[\protect\citeauthoryear{Pounds et al.}{2003b}]{2003MNRAS.346.1025P} 
Pounds K.~A., King A.~R., Page K.~L., O'Brien P.~T., 2003, MNRAS, 346, 1025 


\bibitem{} Power, C., Zubovas, K., Nayakshin, S., King, A.R., 2011,
  MNRAS, submitted


\bibitem[\protect\citeauthoryear{Silk 
\& Nusser}{2010}]{2010ApJ...725..556S} Silk J., Nusser A., 2010, ApJ, 725, 556 


\bibitem[\protect\citeauthoryear{Silk 
\& Rees}{1998}]{1998A&A...331L...1S} Silk J., Rees M.~J., 1998, A\&A, 331, L1 

\bibitem[\protect\citeauthoryear{Sutherland \&
    Dopita}{1993}]{1993ApJS...88..253S} Sutherland R.~S., Dopita
  M.~A., 1993, ApJS, 88, 253

\bibitem[\protect\citeauthoryear{Tremonti, Moustakas, \&
    Diamond-Stanic}{2007}]{2007ApJ...663L..77T} Tremonti C.~A.,
  Moustakas J., Diamond-Stanic A.~M., 2007, ApJ, 663, L77




\end{thebibliography}
\end{document}